\documentclass{article}
\usepackage[square,sort,comma,numbers]{natbib}
\usepackage{amsmath}
\usepackage{color}
\usepackage[]{graphicx}
 \usepackage[final]{nips_2017}
\usepackage{physics}
\usepackage{multirow}
\usepackage{pdfpages}
\usepackage{threeparttable}
\usepackage{inputenc} 

\usepackage[T1]{fontenc}    
\usepackage{hyperref}       
\usepackage{url}            
\usepackage{booktabs}       
\usepackage{amsfonts}       
\usepackage{nicefrac}       
\usepackage{microtype}      

\usepackage{psfig}
\usepackage{csquotes}
\usepackage{array}
\newcolumntype{P}[1]{>{\centering\arraybackslash}p{#1}}

\title{A Chemical Bond-Based Representation of Materials}

\author{
  Van-Doan Nguyen\textsuperscript{(a)}, Le Dinh Khiet\textsuperscript{(a)}, Pham Tien Lam\textsuperscript{(a,b)},  Dam Hieu Chi\textsuperscript{(a,b,c,*)}\\ \\
\textsuperscript{(a)}Japan Advanced Institute of Science and Technology \\ 
1-1 Asahidai, Nomi, Ishikawa 923-1211, Japan.\\
\textsuperscript{(b)}Elements Strategy Initiative Center for Magnetic Materials, \\
National Institute for Materials Science, \\
1-2-1 Sengen, Tsukuba, Ibaraki 305-0047, Japan. \\
\textsuperscript{(c)}Center for Materials Research by Information Integration, \\
Research and Services Division of Materials Data and Integrated System,\\
 National Institute for Materials Science, \\
1-2-1 Sengen, Tsukuba, Ibaraki 305-0047, Japan. \\ \\
Email: \textsuperscript{(*)}dam@jaist.ac.jp
}

\begin{document}

\maketitle

\begin{abstract}

This paper introduces a new representation method that is mainly based on chemical bonds among atoms in materials. Each chemical bond and its surrounded atoms are considered as a unified unit or a local structure that is expected to reflect a part of materials' nature. First, a material is separated into local structures; and then represented as matrices, each of which is computed by using information about the corresponding chemical bond as well as orbital-field matrices of two related atoms. After that, all local structures of the material are utilized by using the statistics point of view. In the experiment, the new method was applied into a materials informatics application that aims at predicting atomization energies using QM7 data set. The results of the experiment show that the new method is more effective than two state-of-the-art representation methods in most of the cases.
 
\end{abstract}

\section{Introduction \label{introduction}}

As remarked in \cite{Lookman2016}, a key element of developing advanced materials is to learn from materials knowledge and available materials data to guide the next experiments or calculations in order to focus on materials with targeted properties. Traditionally, materials knowledge has been discovered by experimental studies. In the last few decades, the knowledge has also been discovered by a conventional approach, called computational materials science, whose scope is to model or predict the behavior of materials based on their composition, micro-structure, process history, and interactions. 

Recently, the development of materials informatics \cite{Agrawal2016,Rodgers2006}, known as a combination of materials science and data science, has opened up a new opportunity for accelerating the discovery of new materials knowledge. Regarding the literature, data science \cite{Dhar2013} is a field of study that employs a wide range of data-driven techniques from a large number of research fields, such as applied mathematics, statistics, computational science, information science, and computer science, in order to understand and analyze data. In materials informatics, data-driven techniques are applied into existing materials data for the purpose of automatically discovering new materials knowledge such as hidden features, hidden chemical and new physical rules, and new patterns \cite{ Ghiringhelli2015,Isayev2015,Yousef2012}. Remarkablely, materials informatics is expected not only to provide foundations for a new paradigm of materials descovery \cite{Rajan2015}, but also to be the next generation of exploring new materials \cite{Takahashi2016}. 

Over the years, a large volume of materials data has been generated \cite{Lookman2015}, and these data are commonly described by using a set of atoms with their coordinates and periodic unit cell vectors and categorized as unstructured data \cite{Lam2017}. In practice, data-driven techniques can be hardly applied directly on materials data. Before applying data-driven techniques, materials data have to be transformed into new representations (or descriptors). The representations need to reflect the nature of materials and the actuating mechanisms of chemical/physical phenomena. In addition, the operators such as comparison and calculations can be performed by using the representations. 

So far, various methods for representing materials have been developed. Behler and co-workers \cite{Behler2011, Eshet2010, Eshet2012} utilized atom-distribution-based symmetry functions to represent the local chemical environment of atoms and employed a multilayer perceptron to map this representation to atomic energy. The arrangement of structural fragments has also been used to represent materials in order to predict the physical properties of molecular and crystalline systems \cite{Pilania2013}. Isayev used the band structure and density of states fingerprint vectors as a representation of materials to visualize material space \cite{Isayev2015}.  Rupps and co-workers developed a representation known as Coulomb matrix for the prediction of atomization energies and formation energies \cite{Faber2015, Matthias2015,Matthias2012}. In \cite{Lam2017}, the authors pointed out that distribution of valence orbitals (electrons) of atoms in materials is important information that should be included in the representation of materials. The author in \cite{Lam2017} also proposed a representation method, called orbital-field matrix, which exploits the distribution.

 
It is well-known that properties of almost materials are determined by the chemical bonds which may result from the electrostatic force of attraction between atoms with opposite charges, or through the sharing of electrons. In addition, chemical bonds hold an enormous amount of energy and building and breaking chemical bonds is part of the energy cycle. Therefore, in this research, we aim at developing a new representation method that mainly based on chemical bonds. In short, the main contributions of the research include (1) a new method to exploit chemical bonds of atoms in materials and (2) a new method to utilize local structures of a material by adopting statistics point of view.

\section{The proposed representation method}

\begin{figure}[t]
    \centering
    \includegraphics[width=0.48\textwidth]{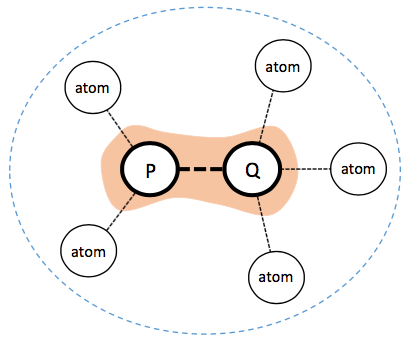}
    \caption{\label{fig:bondkth} Chemical bond $B_k$ and its chemical environment.}
\end{figure}

Generally, a material is composed of chemical bonds that connect atoms together. Let us consider a material, denoted by $X$, which consists of $N$ chemical bonds denoted by $B_1, B_2,..., B_N$. Assume that a chemical bond $B_{k}$ with $1 \leq k \leq N$ is generated by a connection between two atoms $P$ and $Q$, and this bond is surrounded by several other atoms, each of them can connect to atom $P$ or atom $Q$, as illustrated in Figure \ref{fig:bondkth}. The surrounding atoms generate a chemical environment that holds chemical bond $B_k$ in a stable state.
Chemical bond $B_k$ and its chemical environment can be considered as an unified unit corresponding to a local structure of material $X$. In other words, material $X$ can be separated into $N$ local structures corresponding to $N$ chemical bonds and their chemical environments.

Atoms are represented by 32-dimensional vectors, called one-hot vectors  \cite{Lam2017}, which are generated by using a set of valence subshell orbitals $D =  \{s^1, s^2, p^1, p^2, ..., p^6, d^1, d^2, ..., d^{10}, f^1, f^2, ..., f^{14}\}$ (e.g. $p^2$ indicates that the valence $p$ orbital holds 2 electrons in the electron configuration). In addition, we adopt the method of using orbital-field matrix \cite{Lam2017} for representing two atoms $P$ and $Q$. Let $M_p$ and $M_q$ denote two orbital-field matrices corresponding two atoms $P$ and $Q$, respectively. Two matrices $M_p$ and $M_q$ are defined by
\begin{equation}
\begin{split}
	M_p= \vec{P}^T\times \vec{E_p};\\
	M_q= \vec{Q}^T \times \vec{E_q},\\
\end{split}
\end{equation}
where $\vec{P}$ and $\vec{Q}$ are two one-hot vectors corresponding to atoms $P$ and $Q$, and $\vec{E_p}$ and $\vec{E_q}$ are two vectors representing chemical environments of these two atoms \cite{Lam2017}. Two vectors $\vec{E}_p$ and $\vec{E}_q$ are defined by:  
\begin{equation}
\begin{split}
	\vec{E}_{p} =  \sum_{i=1}^{N_p} v_{i} \times \vec{O}_i;\\
	\vec{E}_{q} =  \sum_{i=1}^{N_q} v_{i} \times \vec{O}_i,
\end{split}
\end{equation}
where $N_p$ and $N_q$ are total numbers of atoms connecting to atoms $P$ and $Q$ respectively, and  $v_{i}$ is a coefficient representing the importance role of atom $O_i$.  Weight coefficient $v_{i}$ is defined by
\begin{equation}
	v_i=\frac{\theta_i}{\theta_{max}} \times \frac{1}{r_i^2},
\end{equation}
where $\theta_i$ is the solid angle determined by the face of the Voronoi polyhedron \cite{Aurenhammer1991} that separates atom $O_i$ and its connected atom (atom $P$ or atom $Q$), $\theta_{max}$ is the maximum solid angle among solid angles corresponding the atoms that connect to the connected atom, and $r_i$ is the distance between atom $O_i$ and its connected atom. 

Chemical bond $B_k$ and its chemical environment are then represented by a matrix $U_k$ as follows:
\begin{equation}
	U_{k} =  w_p \times M_p+ w_q \times M_q,
\end{equation}
where $w_p$ and $w_q$ are the coefficients representing the importance roles of atoms $P$ and $Q$ in chemical bond $B_k$, respectively. Coefficients $w_p$  and $w_q$ should be selected according to specific applications; here, we propose that these coefficients are computed by the following equation:
\begin{equation}
	w_p= w_q=\frac{ log_{10}(Z_p \times Z_q)  }{r_{p,q}^n},
\end{equation}
where $Z_p$ and $Z_q$ are the atomic numbers of two atoms $P$ and $Q$ respectively, and $r_{p,q}$ is the distance between these two atoms. 

Because material $X$ contains $N$ chemical bonds, this material is separated into $N$ local structures corresponding to matrices $U_1, U_2,..., U_N$. Regarding the statistics point of view, the set containing the number of local structures, mean and standard deviation of local structures can be used to describe material $X$. Here, mean and standard deviation of local structures, denoted by $\bar{U}$ and $S$, are defined as follows:
\begin{equation}
\begin{split}
	\bar{U} = \{\bar{u}_{i,j}\} \text{\ with \ }
	\bar{u}_{i,j} = \frac{1}{N} \times   \sum_{k=1}^{N} u^{(k)}_{i,j};\\
    S = \{s_{i,j}\}  \text{\ with \ } s_{i,j}=\sqrt{\frac{1}{N} \times  \sum_{k=1}^{N} \abs{ u^{(k)}_{i,j}-\bar{u}_{i,j}}^2};\ \\ 
    \text{\ where \ }
	U_k = \{ u^{(k)}_{i,j} \} \text{ for } k = \overline{1,N} .
\end{split}
\end{equation}
We propose that using this set to represent material $X$. Furthermore, in order to apply data-driven techniques, the representation of material $X$ needs to be transformed into a vector or matrix. Therefore, mean and standard deviation matrices are raved and then combined with the number of chemical bonds in order to form a vector. In other words, material $X$ is represented by a vector as follows:
\begin{equation}
	X=(N,\bar{u}_{1,1},\bar{u}_{1,2},...,\bar{u}_{32,31},\bar{u}_{32,32},s_{1,1},s_{1,2},...,s_{31,32},s_{32,32}).
\end{equation}

Let us consider two materials represented as $X=\{x_i\}$ and $Y=\{y_i\}$ respectively. One can employ various types of distance measurements for measuring the similarity between these two materials, such as listed below:
\begin{itemize}
\item[-]  Euclidean distance \cite{Deza2009}, denoted by $d_{eucl}$: 
\begin{equation}
	d_{eucl}(X, Y) = \sqrt{\sum_{i}(x_{i} - y_{i})^2}.
\end{equation}

\item[-]  Manhattan distance \cite{Krause1987}, denoted by $d_{man}$: 
\begin{equation}
d_{man}(X, Y)  = \sum_{i}|x_{i} - y_{i}|.  
\end{equation}

\item[-] Cosine distance \cite{Singhal2001}, denoted by $d_{cos}$: 
\begin{equation}
d_{cos}(X, Y) = 1 - \frac{\sum_{i}{x_{i} \times y_{i}}}{\sqrt{\sum_{i}x_{i}^2} \times \sqrt{\sum_{i}y_{i}^2}}.  
\end{equation}

\item[-] Bary-Curtis distance \cite{Bray1957}, denoted by $d_{bar}$: 
\begin{equation}
d_{bar}(X, Y) = \frac{\sum_{i}|x_{i} - y_{i}|}{\sum_{i}|x_{i}| + |y_{i}|}.  
\end{equation}

\item[-] Canberra distance \cite{Lance1966}, denoted by $d_{can}$: 
\begin{equation}
d_{can}(X, Y) = \sum_{i}\frac{|x_{i} - z_{i}|}{|x_{i} + y_{i}|}.  
\end{equation}


\end{itemize}

\section{Experiment}
 
To evaluate the new representation method, we applied it into a materials informatics application that aims at predicting atomization energies by using machine learning \cite{Murphy2012}. For analyzing materials data in the application, we selected linear regression technique \cite{Murphy2012} with two learning algorithms, k-nearest neighbors (KNN)  \cite{Murphy2012} and kernel ridge (KR) \cite{Murphy2012}. Additionally, we selected QM7 data set \cite{Matthias2012} for the application. This data set contains 7165 materials (molecules), each of them is composed of a maximum of 23 atoms including C, N, O, S, and H. Coordinates of atoms in materials are presented by Cartesian coordinate system. Information about Coulomb matrix and atomization energies of materials is available in the data set; and the atomization energies are ranging from -800 to -2000 $kcal/mol$. To determine chemical bonds atoms in materials, we employed pymatgen \cite{pymatgen2013}, an open-source library for analyzing materials; however, Voronoi polyhedra \cite{pymatgen2013} could not be determined for 250 materials; thus, they were eliminated from the data set. As a consequence, 6195 materials were actually used in the experiments.

For comparison, we selected two state-of-the-art representastion methods, orbital-field matrix \cite{Lam2017} and Coulomb matrix (eigenspectrum) \cite{Montavon2012,Matthias2012}, as two baselines. For measuring performances of predicting atomization energies we used three well-known assessment methods \cite{Murphy2012}: mean absolute error ($MAE$), root-mean-square error ($RMSE$), and coefficient of determination ($R^2$). Moreover, we applied 5 times 10 folds cross validation into the experiments.

\begin{figure}[t]
    \centering
    \includegraphics[width=0.5\textwidth]{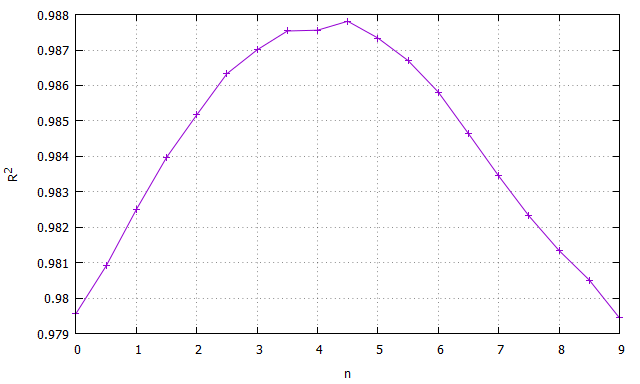}
    \caption{\label{fig:n_R_2} The impact of lengths of chemical bonds on performances of prediction according assessment method $R^2$.}
\end{figure} 

In order to measure the impacts of distances between atoms in chemical bonds (or the lengths of chemical bonds) on performances of predicting atomization energies, we chose KNN learning algorithm with the number of nearest neighbors (denoted by $K$) $K=5$ and Euclidean distance method. The performance according to assessment method $R^2$ was presented in Figure \ref{fig:n_R_2}. As we can see in this figure, the performance increases when $n<5$ and then decreases when $n \geq 5$. It also can be seen that the application archives high accuracy of prediction when the values of $n$ from 3 to 5.    

 \begin{figure}[ht]
    \centering
     \includegraphics[width=0.49\textwidth]{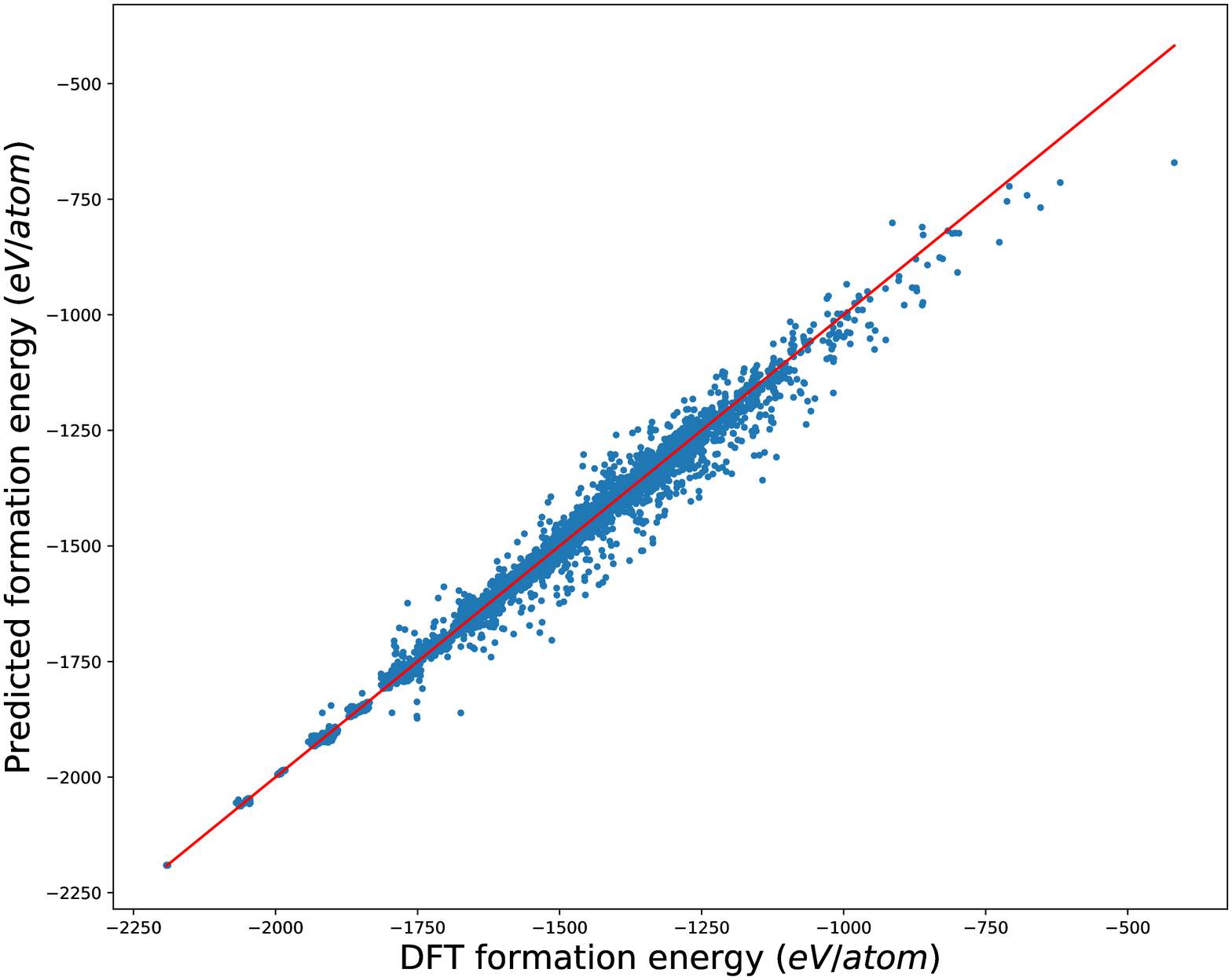} 
        \includegraphics[width=0.49\textwidth]{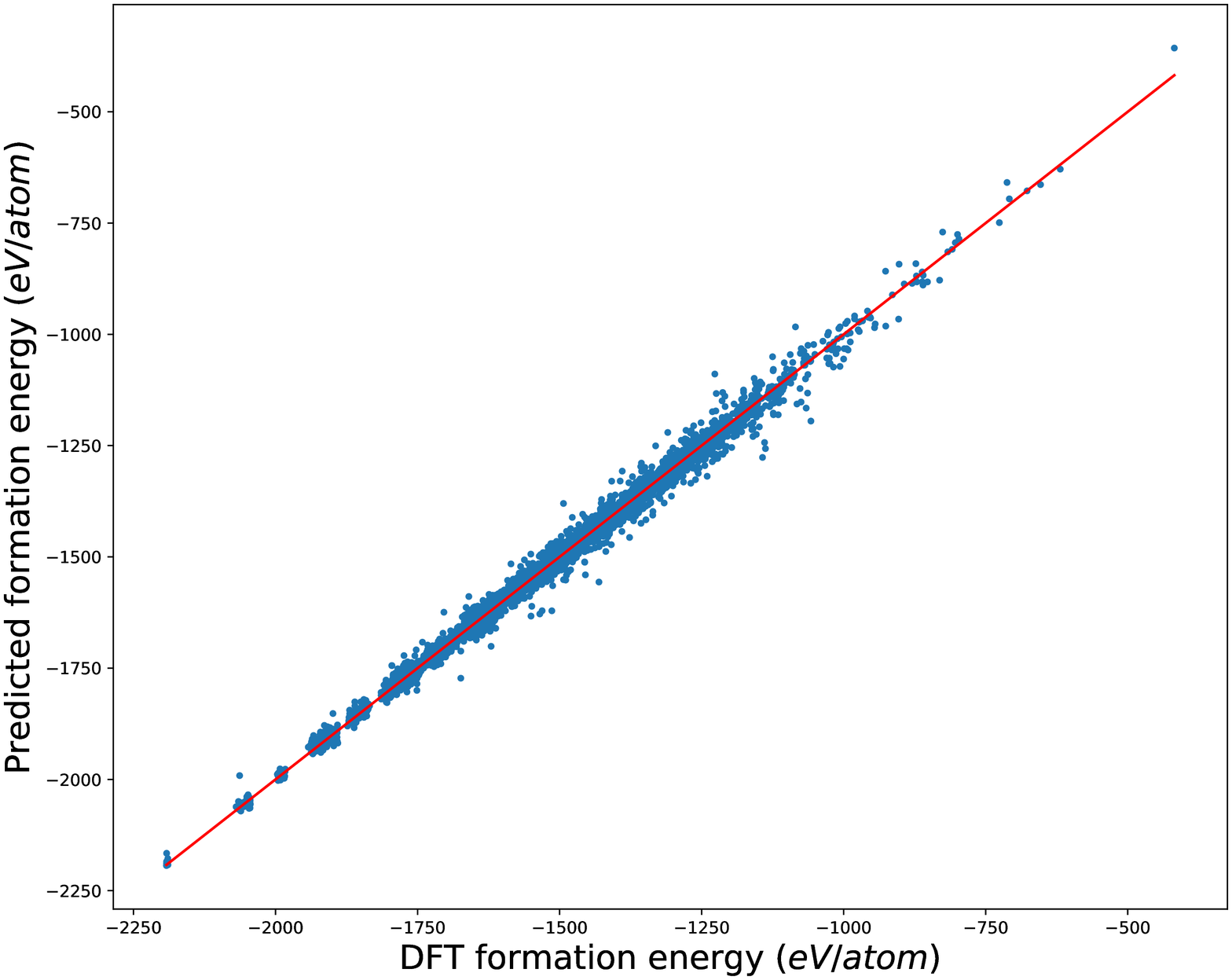} 
      (a)$\ \ \ \ \ \ \ \ \ \ \ \ \ \ \ \ \ \ \ \ \ \ \ \ \ \ \ \ \ \ \ \ \ \ \ \ \ \ \ \ \ \ \ \ \ \ \ \ \ \ \ \ \ \ \ \ \ \ \ \ \ \ \ \ \ \ \ \ \ \ \ \ \ $ (b)
    \caption{\label{fig:knn_krr_chemical_bond} Comparison of predicted atomization energies by using KNN (part a) and KR  (part b) learning algorithms and reference atomization energies calculated by using DFT.}
\end{figure}

Next, we measure the performance of prediction by using the proposed representation method with $n=4$ and both learning algorithms KNN and KR. For KNN, we selected $ K=5$ and Euclidean distance method, and for KR, we selected Laplacian kernel \cite{Murphy2012}. The results of prediction are illustrated in Figure \ref{fig:knn_krr_chemical_bond}. In this figure, parts (a) and (b) show performances of prediction by using KNN and KR learning algorithms, respectively.  As we can observe, the performances according to KR are better than those according KNN.

\begin{table}[ht]
\centering
\small
\caption{Cross-validated $MAE$, $RMSE$ and $R^2$ in the prediction of the atomization energies obtained by using learning algorithm KNN with the selected distance measurement methods.}
\label{tab:KNN}
\def\arraystretch{1.6}
\begin{threeparttable}
\begin{tabular}{c||c|c|c||c|c|c||c|c|c} \hline
Distance 		& \multicolumn{3}{c||}{$MAE$}  & \multicolumn{3}{c||}{$RMSE$} & \multicolumn{3}{c}{$R^2 $}\\ \cline{2-10}
 measure		& (*)  & (**) & (***)& (*)  & (**) & (***) & (*)  & (**) & (***) \\ \hline \hline
Euclidean 	&	\textbf{12.877}	&	14.411	&	78.721	&	\textbf{24.071}	&	30.015	&	102.528	&	\textbf{0.988}	&	0.981   &  0.790	   \\ \hline
Manhattan	&	\textbf{11.447}	&	14.102	&	68.664	& \textbf{22.967}	&	30.218	&	90.181	&	\textbf{0.989}	&	 0.980  &  0.838	  \\ \hline
Cosine			&	\textbf{26.690}	&	42.836	&	85.885  	& \textbf{55.503}	&	97.061	&	111.835&	\textbf{0.934}	&	 0.798  &   0.751	  \\ \hline
Bary-Curtis 	&	\textbf{11.684}	&	14.346	&	68.829		&	\textbf{23.665}	&	30.839	&	90.347	&	\textbf{0.988}	&	0.980   &   0.8372	  \\ \hline
Canberra 		&	 71.527	&	47.010	&	\textbf{18.832}		&	110.528&	72.887	&	\textbf{25.526}	&	0.738	&	0.886   &  \textbf{0.987}	  \\ \hline
\end{tabular} 

\begin{tablenotes}\footnotesize
\item[(*) ]  Chemical bond-based  
\item[(**) ]  Orbital-field matrix 
\item[(***) ]  Coulomb matrix (eigenspectrum)   
\end{tablenotes}
\end{threeparttable}
\end{table}
 
To compare the proposed representation method with two selected baselines, we also selected $n=4$. The results of comparison were summarized in Tables  \ref{tab:KNN} and \ref{tab:KRR}. In these tables, each assessment method for a representation method is represented in a column, and the bold values indicate the best performances in each row and according to the corresponding evaluation assessment method. As detailed in Table \ref{tab:KNN}, the proposed representation method is better than two baselines with the first four distance measure methods, and the representation method by using Coulomb matrix is more effective than the proposed method and the other baseline  according to the Canberra distance method. In addition, it can be seen in Table \ref{tab:KRR}, the proposed method achieves the best performance according to criterion $MAE$, and the representation method by using Coulomb matrix obtains the best performances according to criteria $RMSE$ and $R^2$. However, as observed in Table \ref{tab:KRR}, the performances of the proposed method can be comparable with those of the representation method by using Coulomb matrix.

\begin{table}[ht]
\centering
\caption{Cross-validated $MAE$, $RMSE$ and $R^2$ in the prediction of the atomization energies obtained by learning algorithm KR.}
\label{tab:KRR}
\def\arraystretch{1.6}
\begin{threeparttable}
\begin{tabular}{c||c|c|c||c|c|c||c|c|c} \hline
\multirow{2}{*}{Kernel}		& \multicolumn{3}{c||}{$MAE$}  & \multicolumn{3}{c||}{$RMSE$} & \multicolumn{3}{c}{$R^2 $}\\ \cline{2-10}
 	& (*)  & (**) & (***)& (*)  & (**) & (***) & (*)  & (**) & (***) \\ \hline \hline
Laplacian 	&	 \textbf{9.934}&	13.942	&	9.960	 &	15.106	&24.769	&	\textbf{13.886}	&0.995		&	0.987   & 	 \textbf{0.996}  \\ \hline
 
\end{tabular} 

\begin{tablenotes}\footnotesize
\item[(*) ]  Chemical bond-based  
\item[(**) ]  Orbital-field matrix 
\item[(***) ]  Coulomb matrix (eigenspectrum)   
\end{tablenotes}
\end{threeparttable}
\end{table}

\section{Conclusion}

In this paper, we have proposed a new method for representing materials in materials informatics applications. This method focuses on exploiting information about chemical bonds among atoms in materials and also inherits the benefit of orbital-field matrix representation that is based on the distribution of valence shell electrons. Additionally, we have demonstrated that different similarity measure methods can be integrated with the proposed method. Note that, the proposed method can apply into a large diversity of atomic compositions and structures and facilitate the learning and predicting targeted properties of molecular and crystalline systems.

In the experiment, the proposed method is tested with an application that aims to predict atomization energies; and the results of the experiment indicate that the proposed method is more effective in most the cases when comparing with two selected baselines. In the near future, we plan to further evaluate the proposed method by using different materials data as well as materials informatics applications. 

\bibliographystyle{plain} 
\bibliography{main}
\end{document}